\pdfoutput=1
\documentclass[11pt]{article}

\usepackage{graphicx}
\usepackage{amsmath}
\usepackage{authblk}
\usepackage{hyperref}

\title{Hubble's Law Implies Benford's Law \\
for Distances to Galaxies}
\author{Ronald F. Fox\thanks{School of Physics, Georgia Institute of Technology, Atlanta, GA 30332} \and
Theodore P. Hill\thanks{School of Mathematics, Georgia Institute of Technology, Atlanta GA 30332 and
California Polytechnic State University, San Luis Obispo CA 93407} }
\date{}

\begin{document}

\maketitle
\begin{abstract}
A recent article by Alexopoulos and Leontsinis presented empirical evidence that the first digits of the distances to galaxies are a reasonably good fit to the probabilities predicted by Benford's law, the well known logarithmic statistical distribution of significant digits. The purpose of the present article is to give a theoretical explanation, based on Hubble's law and mathematical properties of Benford's law, why galaxy distances might be expected to follow Benford's law. 
The new galaxy-distance law derived here, which is robust with respect to change of scale and base, to additive and multiplicative computational or observational errors, and to variability of the Hubble constant in both time and space, predicts that conformity to Benford's law will improve as more data on distances to galaxies becomes available.  Conversely, with the logical derivation of this law presented here, the recent empirical observations may be viewed as independent evidence of the validity of Hubble's law.
\end{abstract}

\section{Introduction}
\begin{figure}[htb] 
     \centerline{\includegraphics[width=\textwidth]{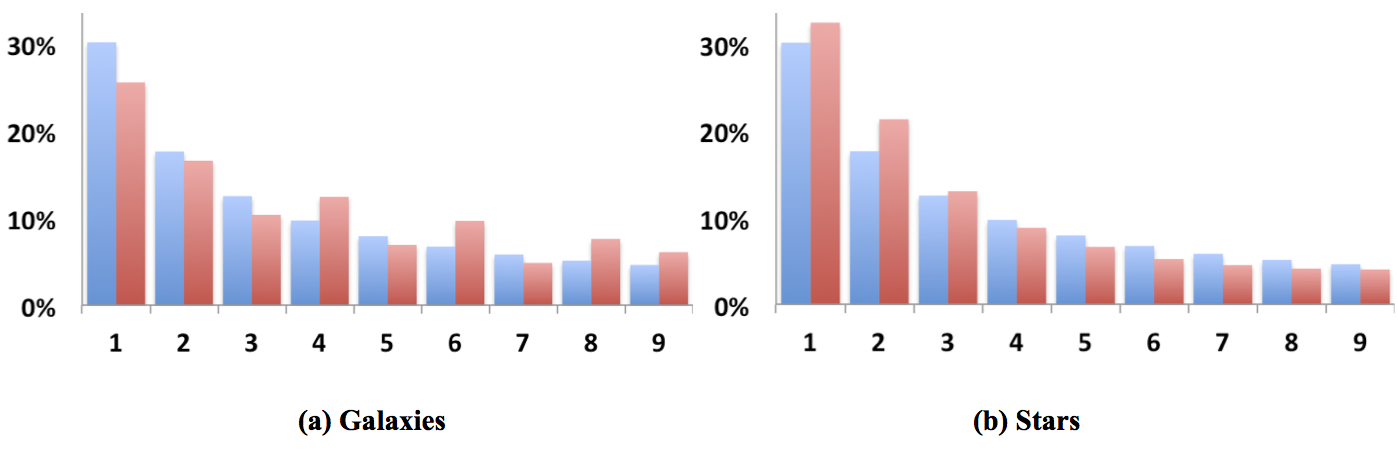}}
                          \caption{Comparison of first digits of (a) 702 galaxy distances (red) with Benford's law (blue); and (b) 115,256 star distances (red) with Benford's law (blue).  Data courtesy of T. Alexopoulos and S. Leontsinis. \label{Fig1}}
\end{figure}

Very recently, Alexopoulos and Leontsinis \cite{AleTL14B} observed that in standard databases of distances to 702 galaxies the first digits are a reasonably good fit to Benford's law;  see Figure \ref{Fig1}(a). 
The main purpose of this note is to show how a Benford distribution of galaxy distances follows from Hubble's law and certain mathematical properties of Benford's law. 
The new galaxy-distance law derived here, which is robust with respect to change of scale and base, to additive and multiplicative computational or observational errors, and to small variability of the Hubble constant in both time and space, predicts that conformity to Benford's law will improve as more data on distances to galaxies becomes available.  Conversely, with the logical derivation of the new galaxy-distance law presented here, the recent empirical observations may be viewed as independent evidence of the validity of Hubble's law.

\vspace{2mm}
\noindent

\section{Hubble's law and galaxy distances}

Let $t_0$ and $t_p$ denote any distant past base time and the present time, respectively, where time is given in (earth) years (for example, $t_p = $00:00:00 hours GMT, 1 January 2015, and $t_0 = t_p - 10^9$).

Let $x(t)$ denote the {\em actual distance} in light years from earth to a generic galaxy at time $t > t_0$, and let  
$\hat{x}(t)$ denote the {\em observed distance} in light years to the galaxy at time $t$. Here ``distance" means the standard {\em proper distance} (as opposed to the {\em co-moving distance}).

Since current measurements are obtained using data from light emitted by the galaxy, the observed distance is the actual distance at the time that light was emitted; in other words, the observed distance satisfies the relation
\begin{equation}\label{eq1}
\hat{x}(t) =  x(t_p - t_c(t)) \, ,
\end{equation}
where $c$ is the speed of light (assumed to be finite and constant), and $t_c(t) = \frac{\hat{x}(t)}{c}$ is the time it took for the observed data to arrive.

Hubble's law in physical cosmology is the observation that distant galaxies are receding from any observation point, such as earth, at a rate that is proportional to their distance away. In terms of the above notation, the idealized Hubble's law (e.g., see \cite{PeaJ99C}) simply states that
\begin{equation}\label{eq2}
\frac{dx}{dt} = Hx \, .
\end{equation}
Assuming $H$ is constant (Hubble's constant), the solution of equation (\ref{eq2}) is
\begin{equation}\label{eq3}
x(t) = x(t_0)\exp(H(t-t_0)) \quad \mbox{\rm for all } t_0 < t \leq t_p \, .
\end{equation}
Together, equations (\ref{eq1}) and (\ref{eq3}) imply
\begin{equation}\label{eq4}
\hat{x}(t_p) = x(t_0)\exp\bigl(H(t_p - t_0 - \frac{\hat{x}(t_p)}{c})\bigr) \, .
\end{equation}
\noindent
Looking backward in time, this implies that the actual distance at the base time $t_0$ in terms of the observed distance at the present time $t_p$ satisfies
\begin{equation}\label{eq5}
x(t_0) = \hat{x}(t_p)\exp\bigl(H \frac{\hat{x}(t_p)}{c}\bigr)\exp(-H(t_p - t_0)) \, .
\end{equation}

\section{Benford's law}

Benford's law is the well-known logarithmic statistical distribution of significant (decimal) digits, dating back to Newcomb \cite{NewS81N}, and popularized by Benford \cite{BenF38L}; the online database \cite{BerHR09B} contains over 800 references to this law. To state it formally, let $S(x)$ denote the decimal {\em significand} (sometimes called {\em coefficient} in floating-point arithmetic) of the positive number $x$; e.g., 
$S(2015) = 2.015 = S(0.02015)$.

With this terminology, a {\em random variable } $X$  {\em is Benford}  if 
$$
Prob(S(X) \leq t) = \log t \quad \mbox{\rm for all } 1 \leq t < 10\, ,
$$
and {\em a sequence} $(x_n) = (x_1,x_2,x_3, \ldots)$ of real numbers {\em is Benford} if
$$
\lim_{N\to \infty} \frac{\#\{n \leq N:S(x_n) \leq t\}}{N} = \log t \quad \mbox{\rm for all } 1 \leq t < 10\, ,
$$
where $\log(t)$ denotes the decimal logarithm of $t$, and $\#\{A\}$ denotes the number of elements in the set {A}.

The obvious relation between these two concepts that will be used below is this:
\begin{equation}\label{eq6}
\begin{split}
\mbox {\rm If } & X_N \, \mbox {\rm is a random variable with values that are equally likely} \\
&  \mbox {\rm to be any of the values } {x_1,x_2,\ldots,x_N}, \mbox {\rm where } (x_n) \, \mbox {\rm is a Benford }\\
&   \mbox {\rm sequence, then } X_N \, \mbox {\rm approaches a Benford distribution as } N\to \infty.
\end{split}
\end{equation}
The most familiar form of Benford's law is the special case of first significant digits, namely
$$
Prob(D_1(X) = d) = \log (1 + \frac{1}{d}) \, , d=1,2,\ldots,9
$$
where $D_1(x)$ is the first significant digit of $x$; e.g., $D_1(2015)$ $ = 2 = $ $D_1(0.02015)$.  Thus, for example, if a dataset (random variable or sequence) is Benford, then exactly $100 \log 2 \cong 30.10\%$ have first significant digit 1, and exactly $100 \log \frac{10}{9} \cong 4.57\%$ have first significant digit 9 (see Figure \ref{Fig1}).

One of the key properties of Benford's law is the fact that it is {\em scale-invariant}, and that it is the only scale-invariant distribution on significant digits \cite[Thm. 3.8]{HilT95B}. In terms of random variables,
\begin{equation}\label{neweq7}
\mbox {\rm If }  X  \, \mbox {\rm is a Benford random variable, then }
 aX \, \mbox {\rm is Benford for all }  a > 0 \, 
\end{equation}
and, in terms of sequences, 
\begin{equation}\label{eq7}
\mbox {\rm If } (x_n) \, \mbox {\rm is a Benford sequence, then } (ax_n) \, 
\mbox {\rm is Benford for all } a > 0.
\end{equation}
Another property of Benford sequences that will play an essential role below is related to the well-known fact that if $b$ is not a rational power of 10, then the sequence $(b^n) = (b, b^2, b^3, \ldots)$  is Benford (e.g., \cite[Lemma 5.3]{BerAH11B}), which follows easily from the uniform distribution characterization of Benford's law and Weyl's theorem about irrational rotations on the circle (see \cite{BerAH11B}). The following generalization of this fact is a crucial part of the argument below; no explicit reference to it is known to the authors, but it is an easy corollary of \cite[Thm. 5.3]{BerABH05O} by taking $\beta_j = \frac{p(j+1)}{p(j)}b$ and $f_j \equiv 0$.
\begin{equation}\label{eq8}
\begin{split}
\mbox {\rm If } & b>0 \, \mbox {\rm is not an exact rational power of 10, then } (p(n)b^n) \, \mbox {\rm is a } \\
& \mbox {\rm Benford sequence for all  non-zero polynomials } p.
\end{split}
\end{equation}
Note that the rate of convergence of the sequence in equation (\ref{eq8}) to Benford depends on both the polynomial $p$ and the base $b$.

\vspace{3mm}
\noindent
{\em N.B.}  None of the familiar classical probability distributions or random variables are Benford, including normal, uniform, exponential, beta, binomial, and gamma distributions.  Similarly, there is no known ``easy'' explanation of Benford's law, and in the context of this note, there is no known easy explanation of why distances to galaxies or stars should be Benford.  In particular, the all-too-common assumption that ``Benford's law applies approximately to any physical quantity that is distributed reasonably smoothly over many orders of magnitude'' is simply wrong.  This can readily be seen by considering a quantity $X = 10^{9} Y$, where $Y$ has a normal (gaussian) distribution with mean 6 and variance 1.  Then $X$ is distributed smoothly over many orders of magnitude, but is far from Benford, since the first digit of $X$ (which is the same as the first digit of $Y$) is 1 with probability less than $1\%$.  For more details on this large-spread fallacy, see \cite{BerAH11BL}.

\section{A Galaxy-distance law}

The main goal of this article is to derive the following law for the distribution of significant digits of distances to galaxies, based on Hubble's law and the mathematical properties of Benford's law stated above. When the value of a variable, such as galaxy distance, is not known {\em a priori}, then a neutral assumption \textendash{}  the simplest and oldest so-called {\em non-informative prior} in statistics \textendash{} is to consider all values in its range equally likely. 

To state a discrete version of the galaxy-distance law concisely, say that a {\em finite lattice} is a collection of regularly-spaced real numbers $\{a, a+\delta, a+2\delta, a+3\delta, \ldots, a+N\delta\}$; for example, the set of numbers $\{100.0, 100.1, 100.2,\ldots, $ $ 499.9, 500.0\}$ is a lattice of 4001 points spaced 0.1 apart starting at 100.0 (i.e., here $a=100.0$, $N=4000$  and $\delta=0.1$). In the framework of this paper, both the starting point $a$ and the spacing $\delta$  are completely arbitrary fixed positive numbers, and only the number of points $N$ varies. The main contribution of this paper is now easy to state.

\vspace{5mm}
\noindent
{\bf Galaxy-distance law}: {\em If the {\bf observed distance} to a galaxy at any given time is equally likely to be any of the values in a finite lattice, then the {\bf actual distance} at that time approaches a Benford distribution as the number of points increases.}
 
\vspace{2mm}
\noindent
{\em Moreover, if the distribution of the significant digits of distances is Benford at any time in the region where Hubble's law holds, then it is Benford at all times in that region.}

\vspace{5mm}
\noindent
To derive this law, fix a generic galaxy and assume that $\hat{x}(t_p)$, the observed distance to that galaxy at the present time, is equally likely to be any of the values $\{a, a+\delta, a+2\delta, a+3\delta, \ldots, a+N\delta\}$  for some $a>0$, $\delta>0$,  and integer  $N \geq 1$. 

Suppose that the observed distance at time $t_p$ is $a + n\delta$ for some $0 \leq n \leq N$;  i.e., $\hat{x}(t_p) = a+n\delta $.
By equation (\ref{eq5}), it follows that $x(t_0)$, the actual distance to the galaxy at the base time $t_0$, satisfies
\begin{equation}\label{eq9}
\begin{split}
x(t_0) = (a + \delta n)\exp(H \frac{\delta n}{c})\exp\bigl(-H(t_p - t_0 - \frac{a}{c})\bigr) = \alpha(a + \delta n)b^n, \\
\quad \mbox {\rm where } \alpha = \exp\bigl(-H(t_p - t_0 - \frac{a}{c})\bigr) > 0 \, \mbox {\rm and } b = \exp(H \frac{\delta}{c}) > 1 \, .
\end{split}
\end{equation}
It may be assumed without loss of generality that $b$ is not a rational power of 10. (This follows since the Lebesgue measure of the set of rational powers of 10 is a null set, that is, has probability zero under any absolutely continuous probability distribution. Alternatively, change $\delta$  by an arbitrarily small amount, if necessary, so that $\log H \frac{\delta}{c}$  is irrational; since the Hubble constant $H$ is not known exactly, this is not an issue.)

By claim (\ref{eq8}), with $p(n)=a+\delta n$, the sequence $((a+\delta n)b^n)$  is Benford, so by the scale-invariance relation for sequences (\ref{eq7}), $(\alpha(a+\delta n)b^n)$  is also a Benford sequence.
By claim (\ref{eq6}) and equation (\ref{eq9}), this implies that if $\hat{x}(t_p)$  is equally likely to be any of the values 
$\{\alpha(a+\delta)b, \alpha(a+2\delta)b^2, \ldots, \alpha(a+N\delta)b^N\}$, then the actual distance $x(t_0)$ at the base time   is a
Benford random variable in the limit as $N\to \infty$.  By equation (\ref{eq3}) and the scale-invariance relation for random variables (\ref{neweq7}), this implies that the actual distance $x(t)$  is Benford at all times $t>t_0$, which completes the argument.  Thus, the above galaxy-distance law predicts that as more data on distances to galaxies become available, the distribution of the significant digits of that data will become even closer to the Benford distribution.

\vspace{2mm}
To relate this law to the empirical data on first digits of distances to galaxies described above, it follows from the Glivenko-Cantelli Theorem, the fundamental theorem of statistics, that a large random sample of galaxy  distances will have an approximately Benford distribution, which is exactly what Alexopoulos and Leontsinis observed (\cite{AleTL14B}; see Figure \ref{Fig1}(a)). The fact that this data is a somewhat less-than-stellar fit to the exact logarithmic distribution of Benford's law may follow from the relatively small data set (702) of galaxy distances available, and/or from truncation of a Benford sequence whose rate of convergence to Benford is unknown (see the remark following claim (\ref{eq8})).

\vspace{2mm}
\noindent
{\em Note}: The above model for galaxy distances was predicated on a snapshot of time (present time) 
where the galaxy distance data  is available (to humans). If, instead, galaxy distances are determined at random times over an astronomically large time period (by definition unavailable to humans), it again follows from Hubble's law that the resulting galaxy distance data will also approach Benford's law, simply because every non-zero exponential function is Benford \cite[Ex. 4.5(iii)]{BerAH11B}. 

\section{Benford distribution of star distances}

Hubble's law is generally applied only to intergalactic distances, i.e., distances of the order of magnitude of megaparsecs or gigaparsecs, which are calculated using completely different methods (such as redshift techniques) than those methods (e.g., parallax) used to calculate star distances.  However, Hubble's law presumably also has a very tiny effect in the much shorter kiloparsec ranges at which star distances are measured, and the above argument applied {\em mutatis mutandis} to the resulting internal expansion of our own galaxy implies that the significant digits of the distances to stars should also approach the Benford distribution. 

\vspace{5mm}
\noindent
{\bf Star-distance law}: {\em If the {\bf observed distance} to a star at any given time is equally likely to be any of the values in a finite lattice, then the {\bf actual distance} at that time approaches a Benford distribution as the number of points increases.}

\vspace{5mm}

In fact, the distances to stars listed in the 2011 HYG database \cite{NasD11H} is an even better fit to Benford's law (see Figure \ref{Fig1}(b)) than the distances to galaxies, perhaps since the sample size (115,256) is so much larger.  Thus, the empirical evidence that star distances are close to Benford's law discovered in \cite{AleTL14B} may be viewed as indirect evidence that galaxies are also expanding at a Hubble-like exponential rate internally, perhaps with a different constant that reflects the gravitational forces involved.

\section{Robustness of the laws}

The Benford distribution is remarkably robust, which perhaps helps explain its widespread ubiquity in empirical data (see \cite{BerHR09B}). With regard to the galaxy- and star-distance laws above, several different aspects of this robustness are relevant.

\vspace{2mm}
\noindent
{\bf (i)} Since the Benford distribution (on significant digits) is scale-invariant (see (\ref{neweq7}) and (\ref{eq7})), the identical galaxy- and star-distance laws hold regardless of what length units are employed \textendash{} exactly the same logarithmic proportions occur with distances given in inches or furlongs as occur with light years. Similarly, since the Benford distribution is the {\em unique} distribution of significant digits that is scale-invariant \cite[Thm. 3.8]{HilT95B}, it follows that if there is any universal statistical distribution at all of the significant digits of galaxy distances, by Hubble's law it must be Benford. 

\vspace{2mm}
\noindent
 {\bf (ii)} Since the Benford distribution is base-invariant as well \cite[Thm. 3.5]{HilT95B}, the analogous galaxy- and star-distance laws also hold with respect to all non-decimal integer bases as well.
 
 \vspace{2mm}
 \noindent
{\bf (iii)} The hypothesis that the possible observed distances are all equally likely can also be relaxed considerably. If the likelihoods of observed distances are decreasing with the distance $x$, say proportional to $1/x$, or are increasing proportional to $a - {1/x}$, then the actual star distances will again be exactly Benford. The likelihood probabilities may even be oscillating, as might be the case when passing through successive clusters of galaxies and intergalactic regions. For example, if the first thousand distance points in the lattice are equally likely but with low probability, the second thousand are also equally likely but with higher probability, and so on alternating in this fashion, then the distribution of actual galaxy distances will again be exactly Benford. These three assertions all require proof; the decreasing case is straightforward using (\ref{eq8}) since every Benford sequence is also ``logarithmic Benford'' \cite{MasBS11s}, and the increasing and oscillating cases follow by analogous but longer arguments, and are beyond the scope of this article. 

\vspace{2mm}
\noindent
{\bf (iv)} The same galaxy- and star-distance laws hold if there are limited (random or deterministic) additive errors in the calculations, since the resulting sequence with errors is also {\em exactly} Benford. This is clear from the following  observation about Benford sequences, an immediate corollary of \cite[Lemma 5.7(i)]{BerAH11B}:
\begin{itemize}
\item[]
      If  $(x_n)$ is Benford and $x_n\to \infty$, then $(x_ n \pm \epsilon_n)$ is Benford for all  $0 \leq \epsilon_n \leq M$, where $M$ is any arbitrary positive number.
\end{itemize}

\vspace{2mm}
\noindent
{\bf (v)} Both laws are also unaltered by independent random multiplicative errors, since the Benford distribution is an attracting distribution in that if $X$ and $Y$ are independent positive random variables, and
either $X$ or $Y$ is Benford, then their product $XY$ is also Benford (\cite[Thm. 6.3]{BerAH11B}).
Thus
\begin{itemize}
\item[] If $X$ is Benford and $E$ is any independent error with $|E| < 1$, then $(1 \pm E)X$ is also Benford.
\end{itemize}

\vspace{2mm}
\noindent
{\bf (vi)} A key part of the above argument involved solution of the differential equation (\ref{eq2}) under the assumption that Hubble's constant $H$ is in fact constant.
Benford's law, however, is also robust in this respect \textendash{} the solution of every differential equation sufficiently close to equation (\ref{eq2}), e.g., one 
where $H$ may vary slightly depending on time or space, is also exactly Benford; this can be seen using  \cite[Thm. 5.3 and Cor. 6.5, respectively]{BerABH05O}.

\vspace{2mm}
\noindent
{\bf (vii)} The above argument is also robust with respect to the {\em magnitudes} of the speed of light and Hubble's constant;  in fact, Hubble's constant could even be negative and the universe contracting.

\section{Conclusions}

Using Hubble's law and mathematical properties of Benford's law, this article derives a galaxy-distance law which predicts a logarithmic distribution of the significant digits of the distances to galaxies, thereby lending theoretical support to recent empirical findings.  The stated galaxy-distance law is robust with respect to change of scale or base, to possible variability of Hubble's constant, and to additive and multiplicative errors in computations. Thus, with the logical derivation of the galaxy-distance law given here, the observations of Alexopoulos and Leontsinis may be viewed as new independent empirical evidence of the validity of Hubble's law.  Similarly, with the analogous logical derivation of the above star-distance law, the close fit of star distances to Benford's law found in \cite{AleTL14B} may be viewed as new empirical evidence that galaxies are also expanding internally at an exponential rate. 

\subsection*{Acknowledgments}

We are grateful to Erika Rogers for bringing the article \cite{AleTL14B} to our attention, to Theodoros Alexopoulos and Stefanos Leontsinis at CERN for providing us with their original data, and to Frederic Rasio for very helpful comments.

\end{document}